# Project overview and update on WEAVE: the next generation wide-field spectroscopy facility for the William Herschel Telescope

Gavin Dalton*.[*a,b], Scott Trager[c], Don Carlos Abrams[d], Piercarlo Bonifacio[f], J. Alfonso L. Aguerri[g], Kevin Middleton[a], Chris Benn[d], Kevin Dee[d], Frédéric Sayède[f], Ian Lewis[b], Johannes Pragt[i], Sergio Picó[d], Nic Walton[j], Juerg Rey[d], Carlos Allende Prieto[g], José Peñate[g], Emilie Lhome[d], Tibor Agócs[i], José Alonso[g], David Terrett[a], Matthew Brock[b], James Gilbert[b], Andy Ridings[d], Isabelle Guinouard[f], Marc Verheijen[c], Ian Tosh[a], Kevin Rogers[a], Iain Steele[e], Remko Stuik[i], Neils Tromp[i], Attila Jasko[k], Jan Kragt[i], Dirk Lesman[i], Chris Mottram[e], Stuart Bates[e], Frank Gribbin[d], Luis Fernando Rodriguez[g], José Miguel Delgado[g], Carlos Martin[d], Diego Cano[d], Ramon Navarro[i], Mike Irwin[j], Jim Lewis[j], Eduardo Gonzales Solares[j], Neil O'Mahony[d], Andrea Bianco[l], Christina Zurita[g], Rik ter Horst[i], Emilio Molinari[m], Marcello Lodi[m], José Guerra[m], Antonella Vallenari[n], Andrea Baruffolo[n].

[a]RALSpace, STFC Rutherford Appleton Laboratory, OX11 0QX, UK; [b]Dept. of Physics, University of Oxford, Keble Road, Oxford, OX1 3RH, UK; [c]Kapteyn Institut, Rijksuniversiteit Groningen, Postbus 800, NL-9700 AV Groningen, Netherlands; [d]Isaac Newton Group, 38700 Santa Cruz de La Palma, Spain; [e]Astrophysics Research Institute, Liverpool John Moores University, IC2 Liverpool Science Park, 146 Brownlow Hill, Liverpool, L3 5RF, UK; [f]GEPI, Observatoire de Paris, Place Jules Janssen, 92195 Meudon, France; [g]Instituto de Astrofisica de Canarias, 38200 La Laguna, TF, Spain; [i]NOVA ASTRON, PO Box 2, 7990 AA, Dwingeloo, Netherlands; [j]Institute of Astronomy, Madingley Road, Cambridge, CB3 0HA, UK; [k]Konkoly Observatory, H-1525 Budapest, P.O.Box 67, Hungary; [l]Osservatorio Astronomico di Brera, INAF, via E. Bianchi 46, 23807 Merate (LC), Italy; [m]Fundación Galileo Galilei, INAF, Rambla José Ana Fernández Pérez, 7, 38712 Breña Baja, TF, Spain; [n]Osservatorio Astronomico di Padova, INAF, Vicolo Osservatorio 5, 35122, Padova, Italy.

## ABSTRACT

We present an overview of and status report on the WEAVE next-generation spectroscopy facility for the William Herschel Telescope (WHT). WEAVE principally targets optical ground-based follow up of upcoming ground-based (LOFAR) and space-based (Gaia) surveys. WEAVE is a multi-object and multi-IFU facility utilizing a new 2-degree prime focus field of view at the WHT, with a buffered pick-and-place positioner system hosting 1000 multi-object (MOS) fibres, 20 integral field units, or a single large IFU for each observation. The fibres are fed to a single spectrograph, with a pair of 8k(spectral) x 6k (spatial) pixel cameras, located within the WHT GHRIL enclosure on the telescope Nasmyth platform, supporting observations at R~5000 over the full 370-1000nm wavelength range in a single exposure, or a high resolution mode with limited coverage in each arm at R~20000. The project is now in the final design and early procurement phase, with commissioning at the telescope expected in 2017.

## 1. INTRODUCTION

With the advent of the new large-area surveys that are under construction from new facilities such as ESA's Gaia satellite and the European Low Frequency Array (LOFAR), WEAVE [1,2] was proposed as a new wide-field spectroscopy facility for the 4.2m William Herschel Telescope (WHT) at the Roque de los Muchachos observatory in the Canary Islands. Although initially developed as a joint project between the UK, Spain and the Netherlands, the project now includes substantial representation from France and Italy.

Gaia[3] will measure precise positions, proper motions and parallaxes for $10^9$ stars with (V ≤ 20) over the whole sky, with medium resolution spectroscopy supplying radial velocity measurements and abundances for the brighter stars in the survey. Extending radial velocity measurements to the depth of the Gaia astrometric survey with radial velocity uncertainties matched to the expected Gaia proper motion errors requires R~5000 spectroscopy from a 4m-class telescope, delivering spectra that will also yield basic stellar parameters and abundance ratios. For the brighter (V~17) stars, extending the follow-up to include higher resolution (R~20000) spectra will provide detailed abundance measurements to allow stars to be labeled by chemical 'fingerprints' that link groups of stars with a common formation history, providing a picture of the mass assembly history of the Milky Way halo.

* gavin.dalton@stfc.ac.uk; Tel +44 1235 446401

The wide-field surveys produced by LOFAR[4] will provide continuum detections of sources identified at 30, 60, 120 and 200 Mhz, the majority of which are expected to be extragalactic sources dominated by radio emission from star formation activity, but with no redshift information. At the expected detection limits, LOFAR should be sensitive to star formation rates $\geq 10 M_\odot$/yr out to substantial cosmological distances, implying emission line fluxes for H-$\alpha$ and O[II] that should be detectable in a 1 hour exposure with 4m-class instrumentation. Spectroscopic identification of a substantial fraction of these sources will yield metallicities and velocity dispersions, and hence stellar populations and star formation histories. In parallel, follow-up of LOFAR surveys will provide unbiased samples for cosmological investigations.

At lower redshifts, the study of galaxy evolution calls for spatially resolved spectroscopy to reveal details of the internal structure and kinematics of the stellar populations. A high multiplex fibre-fed spectrograph can easily be extended to accommodate fibre-based integral field units to allow such observations to be extended to large samples of galaxies over a moderate redshift range, provided these can be accommodated within the focal plane architecture of the instrument.

The WEAVE facility is now well into its final design phase, with first light currently scheduled for mid-2017. In this paper we will summarize the current state of the design with reference to more detailed descriptions of specific aspects of the instrument that are given elsewhere in these proceedings[5,6,7,8,9,10,11,12]. In Section 2 we give a general overview of the development of the WEAVE facility in the context of the existing configuration of the WHT. Section 3 describes the new 2 degree field prime focus corrector. The adopted fibre positioning architecture is detailed in Section 4, together with a summary of recent developments with a prototype in the laboratory. The positioner is linked to the WEAVE spectrograph by 32m of fibre cable, which is outlined in Section 5, followed by a description of the spectrograph itself in Section 6. WEAVE is expected to operate initially for a 5-year survey campaign, for which all data will be automatically processed through to standard analyses of the object spectra. Section 7 outlines the architecture of this data-flow system through to the WEAVE archive. We summarize the expected performance and the instrument schedule in Section 8.

## 2.  THE WEAVE FACILITY

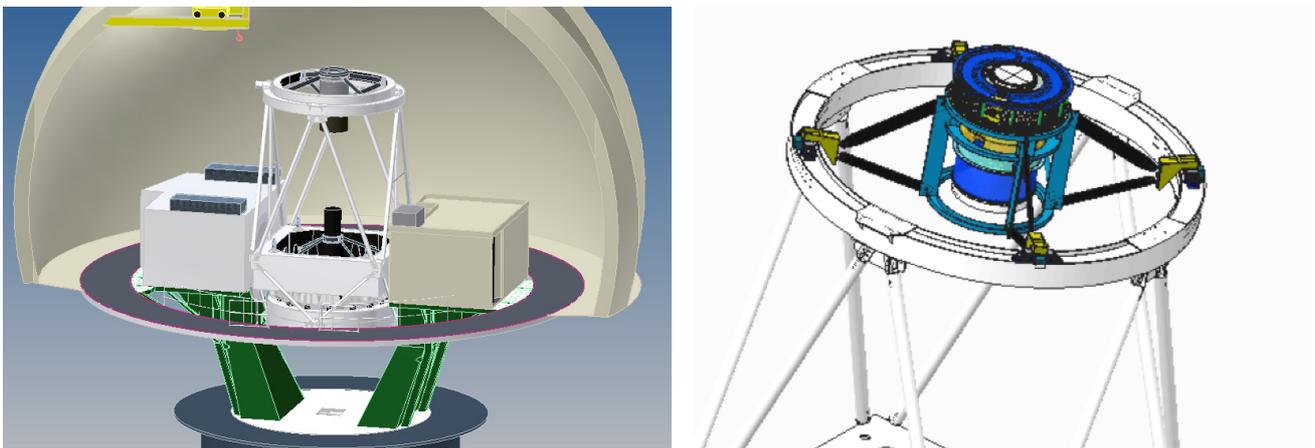

Figure 1: (*left*) The current layout of the WHT showing the two Nasmyth enclosures, GRACE (left) and GHRIL (right). The current top-end assembly is housed in the inner flip-ring. (*right*) WEAVE top-end structure with the current flip-ring removed and the new vanes mounted to focusing units on the outer ring. The turntable assembly shown at the top of the structure provides a general purpose instrument mount designed to accommodate the WEAVE fibre positioner or other future instrumentation.

In its current configuration, the WHT provides a flexible platform for diverse instrumentation, with provision for instruments at Cassegrain, prime-focus, folded-Cassegrain and two stabilized enclosures at the Nasmyth foci. From the outset it was clear that the new top-end package required for WEAVE could not be accommodated within the existing top-end ring structure, although the total mass of the current flip-ring and centre-section of 5.66T provides sufficient margin for the new system while maintaining the current mass and balance budget for the telescope. The original concept for WEAVE[2] (Figure 1, left) was to replace the current flip-ring with a new inner ring structure that would support the WEAVE top-end and maintain the pre-tensioning of the support vanes in a structure that could easily be deployed and removed. The initial concept for the new prime focus corrector relied on a translation of the last corrector element to achieve focus control, but this approach led to rapid changes in the focal plane image scale[5] that could not be accommodated by the positioner. A revised approach allowing for focus control by axial translation of the whole ring structure was then devised, but detailed analysis[6] subsequently shows that in this configuration the inner ring is under very low stress and could ultimately be dispensed with altogether. In the final configuration (Figure 1, right), a handling frame is used to retain the pretension in the vanes during installation and storage. The total mass of the deployable system, including the fibre positioner is around 4T. This approach has the added advantage of reducing the space required for the WEAVE top end assembly when off the telescope. The top-end assembly is adjusted for focus and tilt by means of off-centred cylindrical cam actuators which push on 4mm thick titanium blade springs to which the outer ends of the support vanes are anchored. With this arrangement, the focus position can be adjusted by ±3mm with an accuracy of 5μm, compared to the 20μm absolute positioning error required to maintain optimal focus as the temperature of the structure changes.

The prime focus corrector assembly mounts directly to the inner side of the central support structure. The field rotator bearing is mounted directly to the upper end of this structure, such that the last corrector lens passes through the rotator cable wrap. For alignment purposes, the optical axis from the corrector must be coincident with the rotation axis of the field rotator to within 60μm, although the precision with which this offset must be known is around twice this value. The field rotator is specified to operate over a 120 degree range with < 0.05"/hour tracking error.

The provision of temperature-controlled enclosures on both WHT Nasmyth platforms provides an optimal location to site the WEAVE spectrograph (in GHRIL) and the control electronics for the fibre positioner (in GRACE). When WEAVE is operational, it will still be possible to replace the original top end flip-ring and return to more general operations with support for a subset of the existing instrument suite, or visiting instrumentation, either at Cassegrain or in GRACE, but the GHRIL enclosure will be fully occupied by the WEAVE spectrograph. When WEAVE is 'off' the telescope, the top end assembly can be stored in the mirror coating area of the telescope building, where it will be possible to operate the positioner for calibration and maintenance.

## 3. PRIME FOCUS CORRECTOR

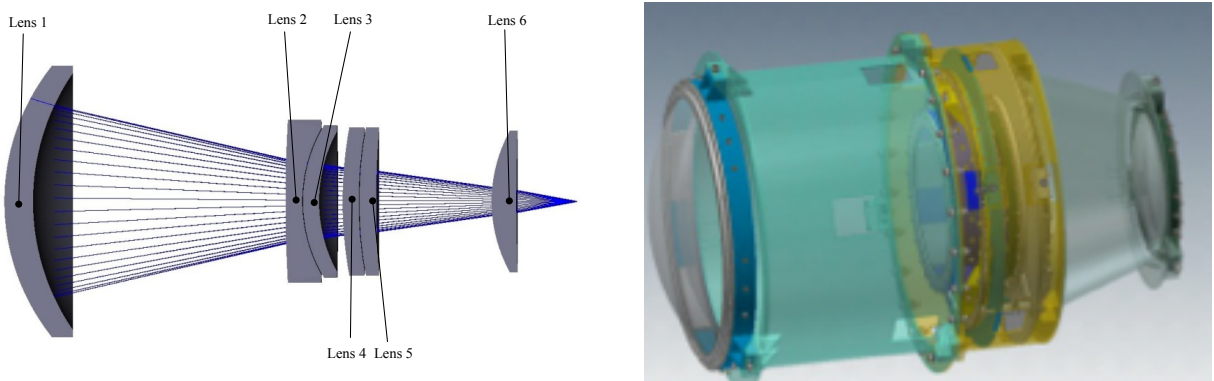

Figure 2: (*left*) The optical layout of the corrector. (*right*) Corrector housing showing the ADC assembly (yellow) and the spacing structure to the lens 1 invar cell.

The prime focus corrector[5] passed its optical final design review in July 2013 and is now in manufacture, with the delivery of the completed lenses expected early in 2016. The basic design began as a 4-element lens system with contacted prismatic doublets for the central two 600mm diameter elements to form a counter-rotating atmospheric dispersion compensator (ADC). In the final design phase, it was decided to separate the prismatic doublets, both to provide additional surfaces for control of the wave front that would allow the severity of the aspheric surfaces in the corrector to be relaxed, and to avoid possible issues with CTE matching within these large lenses. The lenses are Silica, N-BK7, PBL1Y, PBL1Y, N-BK7, Silica (Figure 2). The first lens is 1100mm in diameter and will be slumped at Corning before shipping to the polisher. Due to the large size and curvature of this element, we have elected to leave it uncoated on the grounds of risk. Considerable attention has been given to the mounting of these large lenses, and the current design is based on the successful approach used for DECam[13]. After detailed tolerancing of the manufacture of the lenses and the assembled corrector the expected image quality is < 1" (80% EED, <0.66" FWHM). When combined with the expected 75[th] percentile of the seeing distribution at the WHT this implies a recovered PSF of FHWM < 1.15" for stellar sources.

# 4. FIBRE POSITIONER

WEAVE employs a fibre positioner concept[7] developed directly from the successful pick and place design used in the 2dF instrument at the Anglo Australian Telescope[14]. This architecture was chosen on the basis of the low development risk and the overall cost of the instrument, as well as the relative ease with which integral field units could be implemented within the architecture of the multi-object (MOS) mode of the instrument. The final concept uses off-the-shelf gantry systems with an integrated PLC control system that implies minimal development for the low-level control system. Like 2dF, WEAVE uses a tumbler mechanism and two sets of fibres to allow one target field to be observed at the telescope focus while the next field is being prepared. This allows sufficient time for a large number of fibre moves within the expected typical exposure time, and provides a minimal idle time overhead to change from one target field to the next.

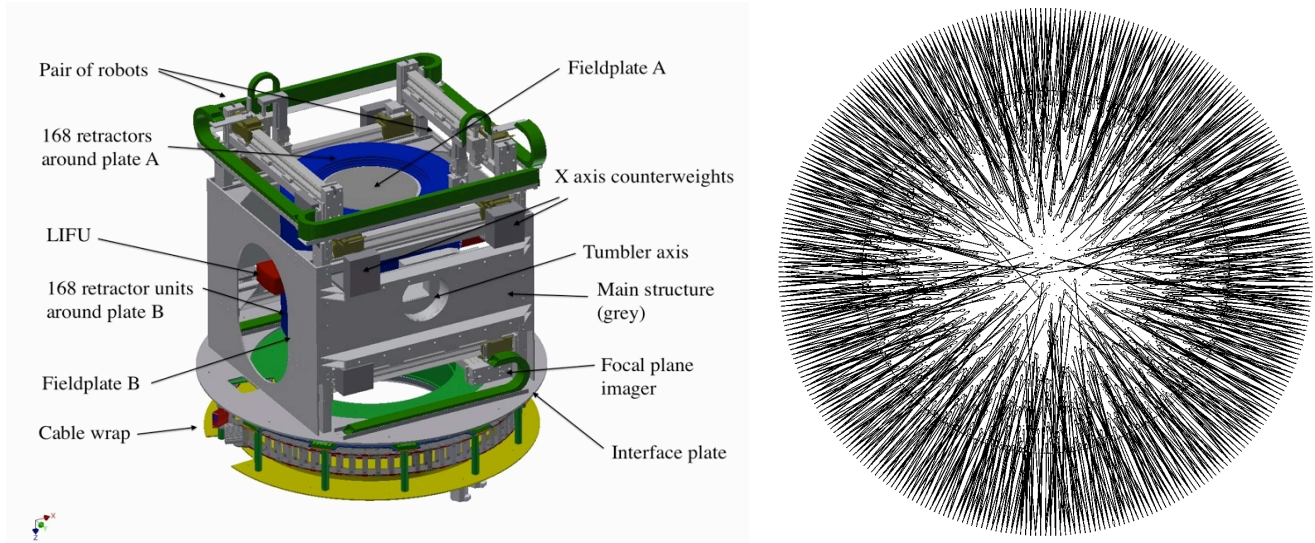

Figure 3: (*left*) The general layout of the fibre positioner system. (*right*) Example of a fully configured field. The black circle marks the optical field of view.

As shown in Figure 3, the positioner makes use of a double gantry system, designed such that each robot will be able to reach around 70% of the fibre park positions. The use of two robots relaxes the time (derived from detailed simulations[8]) per move from <2s to just under 4s in the typical case where a full complement of 1000 fibres is to be moved during a 1 hour reconfiguration. A third gantry system provides a deployable imager, co-focal with the fibres, that can be moved to any location within the full 2° field of view. Each gantry is equipped with a field-plate imaging system to allow in-situ calibration between the coordinate systems of the gantry and the field plates after plate exchange.

Field changeover is achieved by parking all three gantries, then rotating the central tumbler assembly, which houses the field plates and all the fibre retractor units, about its axis. The geometry of the tumbler (length x outer diameter) is set by the retractor units: these are set 70mm outside the useful 205mm radius of the optical field, and push the tumbler diameter to a total of 940mm. The starting point is set by the need to arrange 1000 fibres around the field circumference, and the end point by the limiting bend radius applied to each of three stacked pulley systems within each unit. Each group of three fibres have tiered park locations at the field periphery. The length of the tumbler is determined by the diameter of the tumbler axis and the length required for the inner pulley set in the retractor to allow its fibre to be deployed just beyond the field centre. The outer fibres are restricted to units of the same length, and so are limited to an outer annulus of the science field. This implies that the central 25% of the field is only accessible by 33% of the fibres, which is not expected to present significant difficulties for the majority of science fields. With this geometry, setting the gantry park positions to be clear of the volume of rotation of the tumbler sets the overall size of the positioner, with the outer diagonal of this structure setting the central obstruction diameter for the telescope in the WEAVE configuration at 1.8m.

The 20 mini-IFUs are mounted within the 10 of the retractor slots on one of the field plates, thus reducing the total count of MOS fibres to 940. The mini-IFUs, each comprised of a close-packed array of 37 fibres with the same 1.3" fibre aperture as the MOS mode, will use a scaled-up version of the MOS retractor, but with only two bundles per unit. Each field plate will also be equipped with 8 dedicated fibres for acquisition and auto-guiding. These will be coherent imaging bundles, as currently used in AF2 at the WHT, with a field of view of ~3.5". With 8 guide stars per field, WEAVE aims to correct for the bulk effect of differential refraction over the observation by tracking the position of each star across its target bundle.

The provision of a large monolithic IFU, utilizing the full slit length of the spectrograph to give a field of 1.3'x1.5' field comprised of 540 close-packed 2.6" aperture fibres, is achieved by mounting the IFU at the 90° (halfway) position of the tumbler assembly. The primary objective for the large IFU is to survey the low-surface-brightness extremities of low redshift galaxies at moderately high (R~10000) spectral resolution, and hence the large apertures allow a maximal field of view for a contiguous field with the available slit length. These observations imply long (~8 hour) exposures, for which sub-spaxel repeatability is highly desirable. The large IFU is therefore accompanied by its own dedicated autoguider camera to avoid possible flexure issues between the IFU and the sky imaging gantry system as there is no means to provide a referencing grid for the imager gantry at the IFU location. The large IFU will be counterbalanced by the main autoguider head that views the output of the imaging bundles from the two field plates.

## 5. FIBRE SYSTEM

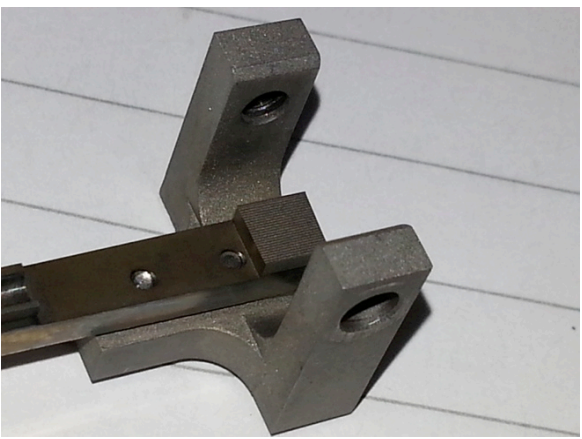

Figure 4: example of a v-groove array slit block manufactured by wire erosion at Oxford.

The WEAVE science fibres[11] (nearly 3200 in all) exit the tumbler through its principal axis and are routed through a cable wrap on the instrument rotator to pass along the prime focus support vanes. They are then routed down the telescope truss, over the elevation cable wrap and into GHRIL to end in slit units for the spectrograph. Including the cable wrap and spare fibre stored within the retractor units to facilitate repairs at the positioner end, the total fibre length has been estimated at 32m. Within the retractors, the 6 MOS fibres are each clad in a protective PEEK tubing, 0.5mm outer diameter, until they merge into a single polyurethane tube of cross-section 2.7x4mm. Each group of four retractors then feeds into a further junction to give 24 fibres in the same polyurethane tubing which passes through the prime cable wrap, down the telescope and over the elevation cable wrap inside a ruggedized conduit. Inside the GHRIL room, this bundle of 24 fibres emerges into a smaller outer PEEK tube of cross-section 1.2x1.7mm which feeds into a cable clamp on the back of a 24-fibre sublit block (Figure 4). The sublit blocks are angled to approximate a curve in two dimensions to form the input slit of the spectrograph, with the angle being common to each group of four blocks. An end-end prototype of one 24-fibre cable from fibre-buttons to sublit has been fabricated by SEDI-ATI. Tests of various aspects of this bundle are currently underway[11].

# 6. SPECTROGRAPH DESIGN

The spectrograph design[9] passed its optical FDR in January 2014 with manufacture of the optics commencing shortly. The optical layout is shown in Figure 5.

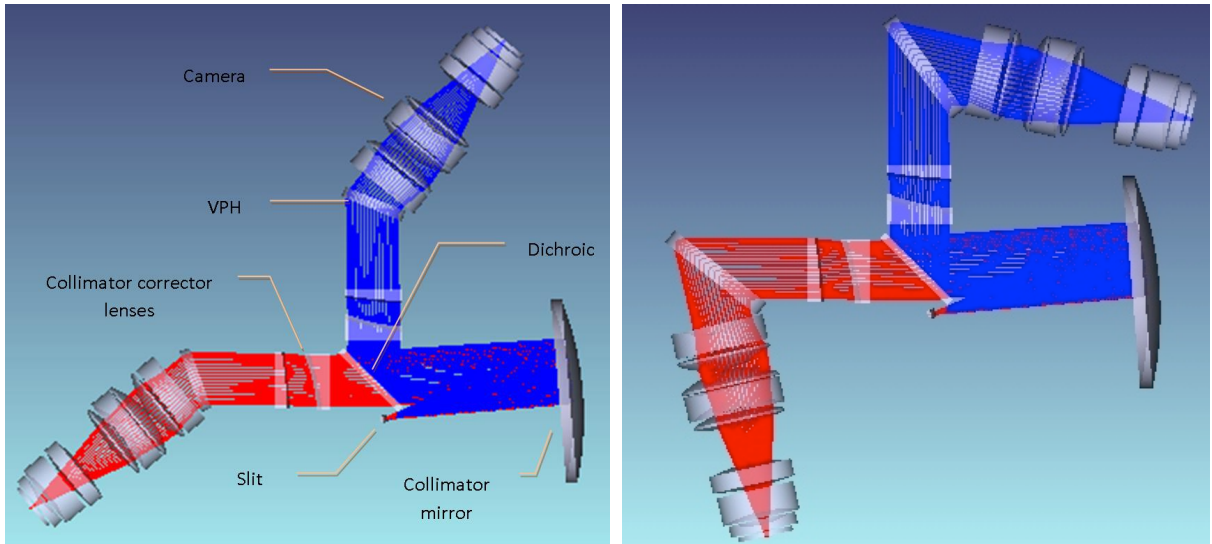

Figure 5 WEAVE optical layout, low (left) and high (right) resolution mode.

The spectrograph consists of a reflective f/3.1 collimator with two sets of field corrector lenses (spherical surfaces) behind the 600nm dichroic, with an 8-lens transmissive camera in each arm, designed such that the lens surfaces, including a single asphere, are identical in the two cameras, with only the coatings and lens-spacings changing. As shown in Figure 5, the spectrograph is switchable between low- and high-resolution modes (R=5000,20000) by a change of grating and a rotation of the cameras. The beam diameter is 190mm and the camera focal ratio is f/1.8. The original concept assumed a pair of 3kx8k e2V CCD231-68 detectors in each arm, as in the LBT MODS spectrograph[15], with a gap in the spatial direction. However, the design has now evolved to accommodate a pair of CCD231-C6 6kx6k detectors in each camera. This change arises from a number of considerations, including spectral curvature on the chip, the need to suppress the recombination ghost from the VPH gratings, and the better overall performance and wider availability (and hence lower cost) of the C6 devices. With this configuration, the nominal focal plane format remains at 6k pixels in the spatial direction and 8k pixels in the spectral direction, such that it is possible to locate the gaps in the two arms at regions of minimal impact on the science.

With this configuration, the 1.3" input fibres give a spectral coverage of 366nm-606nm in the blue arm and 579nm-959nm in the red arm at a typical resolution of 5000, or 404-465nm/473-545nm/595-685nm at a resolution of 20000. There is, in principle, a possible location for a 4[th] high resolution grating in the red arm, but this is not included in the project specifications. The overall spectrograph mechanical concept is shown in Figure 6.

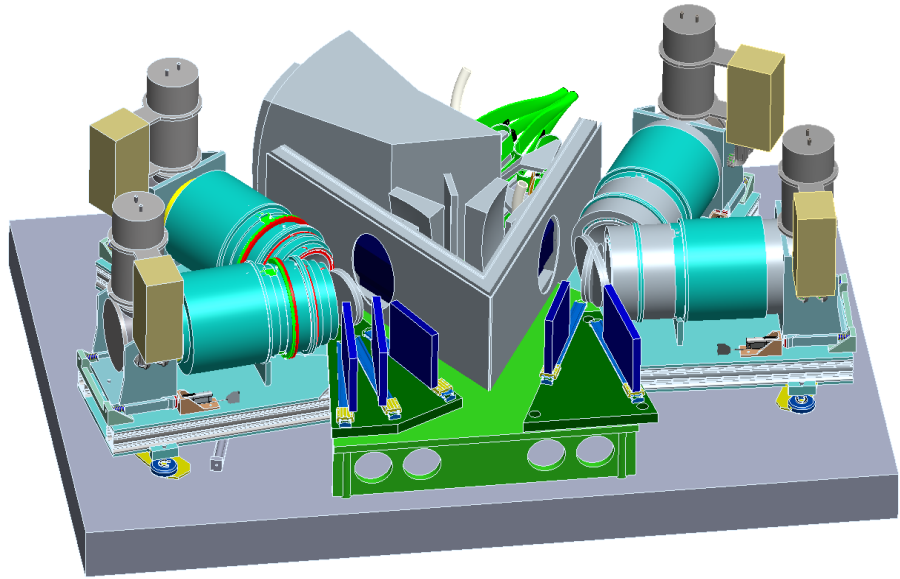

Figure 6 Overview of mechanical design with the cameras shown overlaid in both low- and high-resolution positions.

The fibre slits are mounted in a cross-configuration as shown in Figure 7: The four back-illumination units provide a light source for the fibres on the plate that is being configured so that these can be imaged from the gripper-mounted cameras in the positioner. The central position in the cross defined by the four back-illumination units is the active slit location, and each slit can be driven forwards within the mechanism to a kinematic seat corresponding to the optical entrance of the spectrograph. When the top end unit is dismounted from the telescope, the individual fibre slits can be withdrawn from the exchange mechanism, mounted in protective covers, and then passed out of the GHRIL enclosure. The fibres can then be coiled around drums mounted on the top-end assembly handling frame for safe stowage.

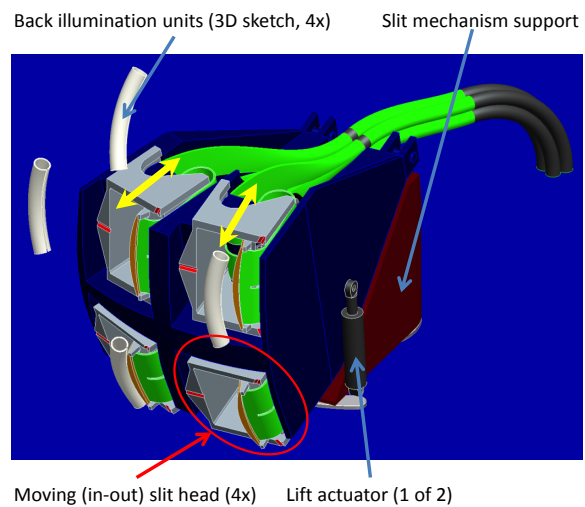

Figure 7 Detailed view of the slit exchange mechanism.

## 7. WEAVE DATA FLOW

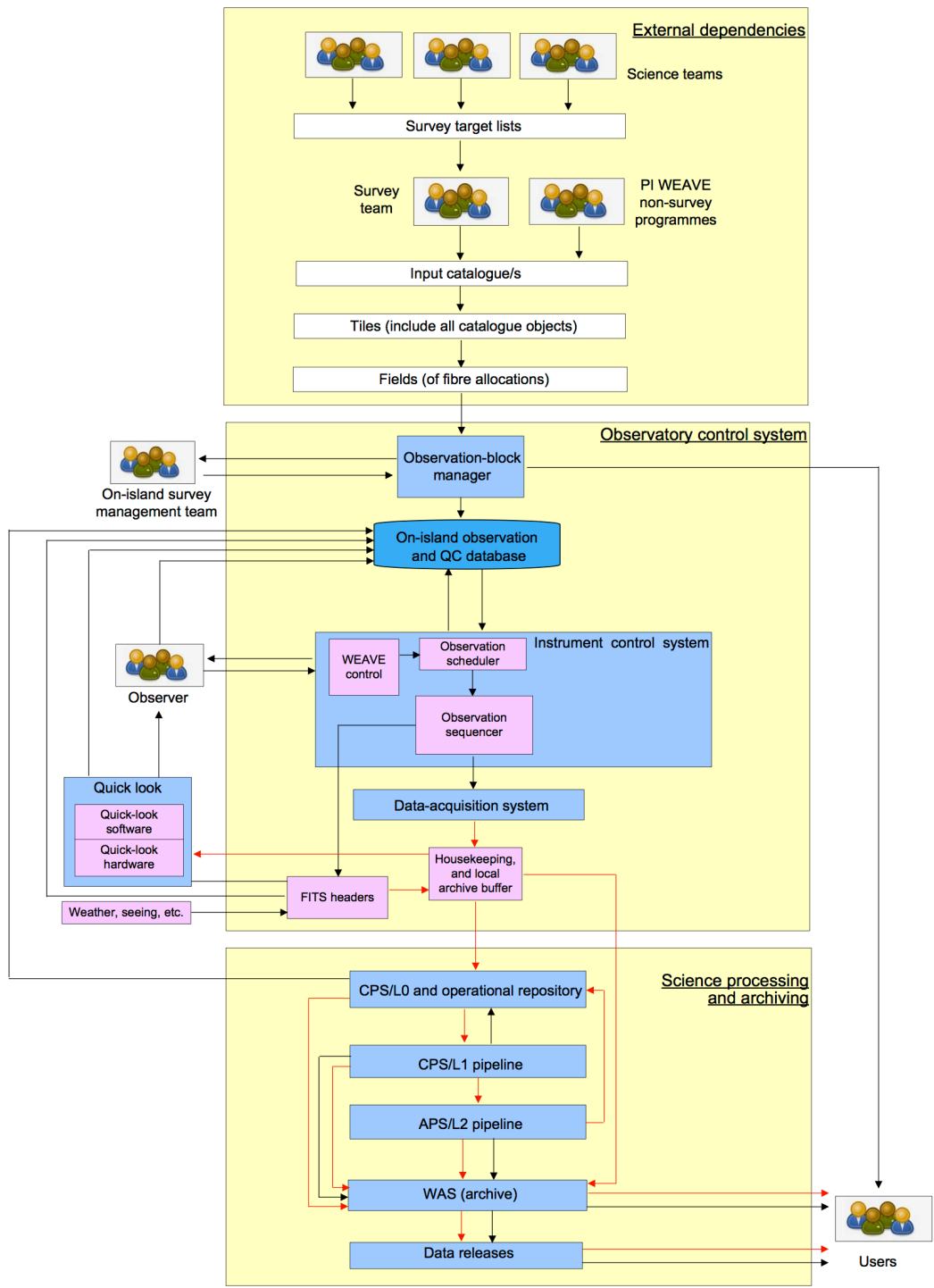

Figure 8 Schematic of the overall data flow.

WEAVE is expected to operate in full survey mode when on the telescope, with the scientific targets for each of the various programmes merged to produce common catalogues for observations in the same regions of sky with the same exposure times. Broadly speaking, these reduce to a combined low resolution survey of the extra-Galactic sky and Galactic Halo; a combined high resolution survey of the same region with chemodynamics targets merged with bright quasars for a Lyman-alpha forest survey; low- and high-resolution surveys of the Galactic disk and clusters; and a number of smaller surveys including galaxy cluster surveys, deep galaxy evolution surveys; IFU surveys of nearby galaxies and smaller PI-driven programmes. Each of these operational surveys will produce target lists and tile centres which will then be used to determine fibre configurations and observation blocks (OBs). A central database of OBs will be maintained at ING. Night-time observations will be selected from this database to produce an observing plan for each night. Science data frames will be passed through an automated quick-look reduction pipeline at the telescope, to return quality control information to the operator and flag possible instrument failures. All data will be passed to CASU in near-real time, and processed daily by the WEAVE Core Processing (CPS) pipeline[12]. The output of the CPS, comprising reduced, calibrated and combined 1-D spectra, will be passed to the WEAVE Advanced Processing (APS) pipeline at the IAC, which will determine classifications, redshifts/radial velocities, along with stellar parameters and basic abundance information. The outputs from both CPS and APS will be stored, along with all raw data frames in the WEAVE archive, from which regular phased releases of the spectra and advanced products will be made available to the WEAVE community.

## 8. OVERALL PERFORMANCE

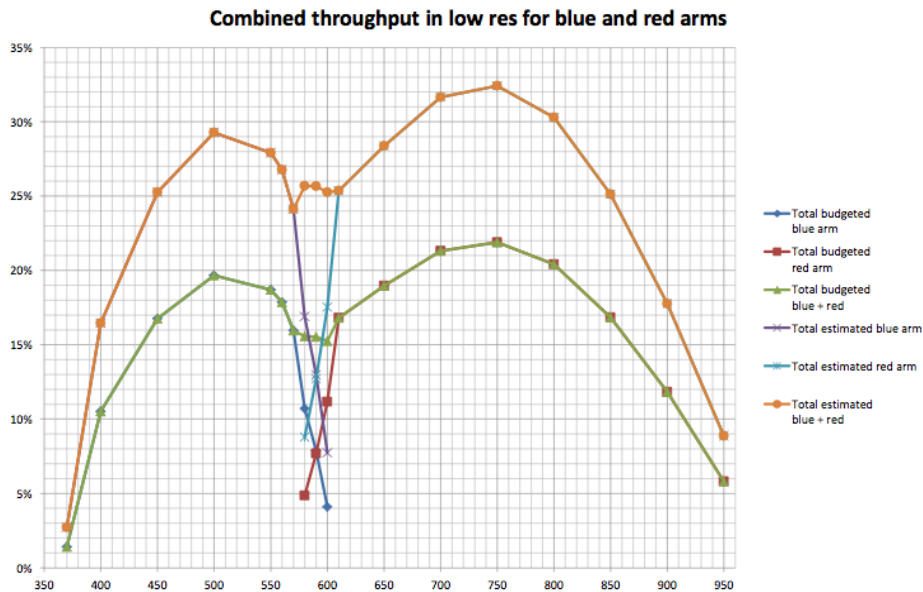

Figure 9 Total system performance for the low resolution mode, including all elements from the bottom of the atmosphere to the detector readout.

The total estimated throughput of WEAVE is shown in Figure 9, together with a more detailed breakdown and predictions for the high-resolution modes in Figure 10. Figure 9 also shows the budgeted system performance developed from the WEAVE science requirements. The expected performance in low-resolution mode is above 25% from 450-850nm.

In Figure 11 we show the expected radial velocity accuracy that should be achievable from the WEAVE spectra compared to the expected radial velocity and photometric limits from Gaia.

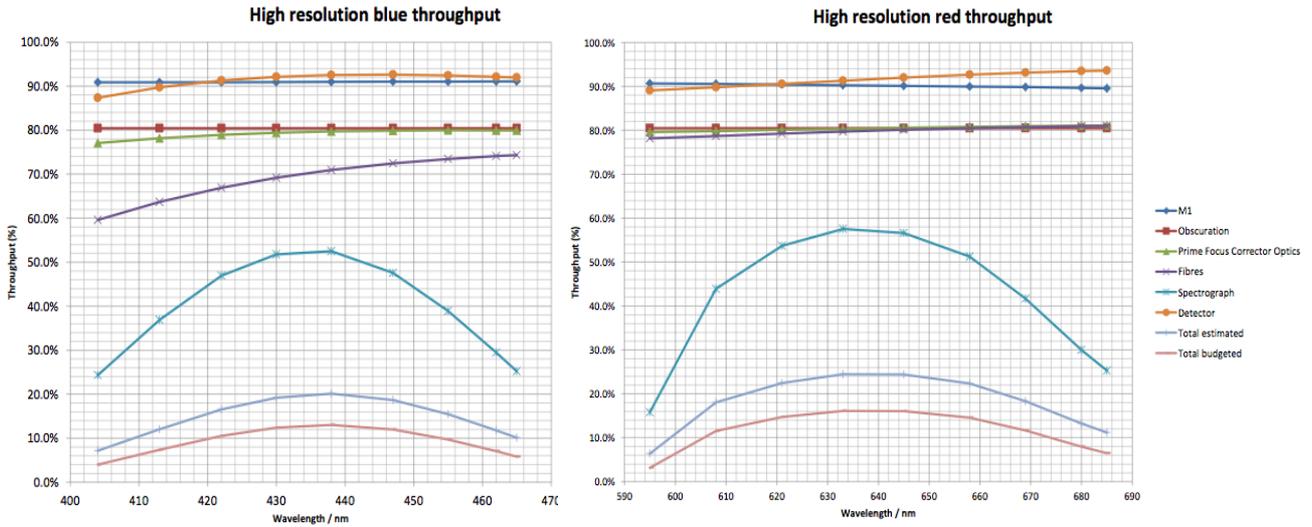

Figure 10 Total system performance for the high-resolution mode, showing contributions to the overall performance from the different systems.

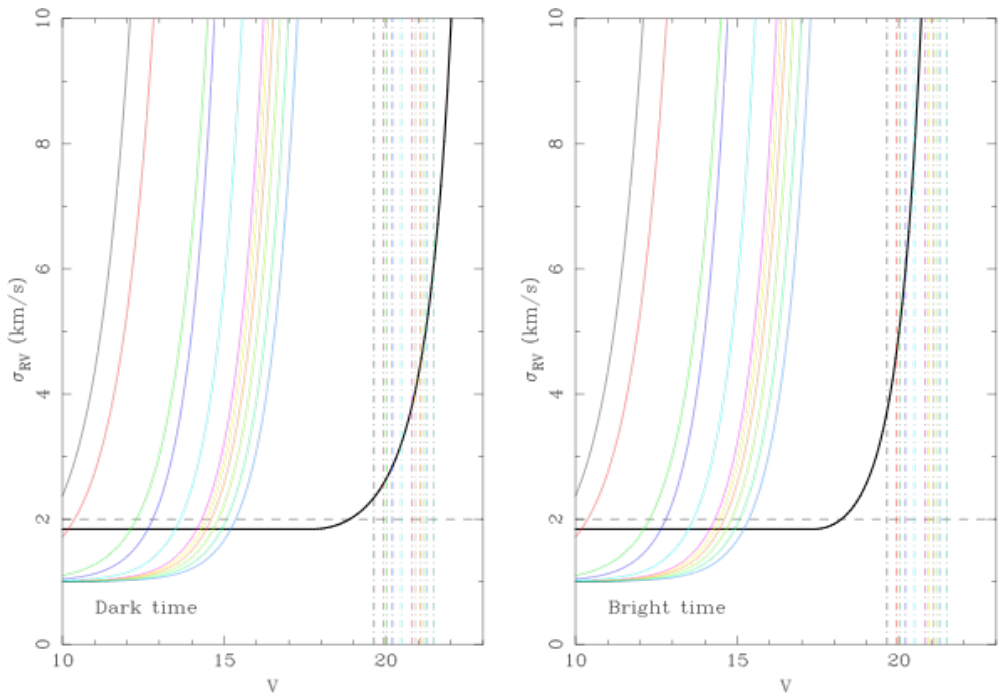

Figure 11 Expected radial velocity accuracies recovered from WEAVE observations (black lines) in dark and bright time for a 1 hour exposure on stellar sources in 0.8" (75 percentile) seeing. Solid colored lines show predictions for the expected performance of the Gaia RVS[3] for stars of different spectral types, while dashed lines show the Gaia photometric limits.

# 9. ACKNOWLEDGEMENTS

The WEAVE project is supported through the Isaac Newton Group Partnership and by grants from the UK Science and Technology Facilities Council (STFC), the Nederlandse Organisatie voor Wetenschappelijk Onderzoek (NWO), the Nederlandse Onderzoekschool voor Astronomie (NOVA), the Instituto de Astrofísica de Canarias (IAC), the Région Île de France and the Instituto Nazionale di Astrofisica (INAF), as well as by further in-kind contributions from the Centre National de la Recherche Scientifique (CNRS), and Konkoly Observatory. We would also like to thank René Rutten, Roland Bacon, Axel Yanes, Eli Atad, Paul Jolley, Ian Parry, Peter Doel, Jorge Sanchez, Gary Hill, Paolo Spano, Vanessa Hill, Olivier Schnurr, Barry Fell and Nigel Morris for their invaluable assistance in various project reviews.